\documentclass[12pt,preprint,a4paper]{article}

\usepackage{amsmath, amssymb, mathrsfs, graphicx, subcaption, dsfont, enumerate, epstopdf, xcolor}
\usepackage{cite} 
\usepackage[colorlinks, allcolors=blue!70!black, linktocpage]{hyperref}
\usepackage[inline]{enumitem}
\usepackage[utf8]{inputenc}

\numberwithin{equation}{section}

\def\half{\frac{1}{2}}
\def\eq#1 { \begin{equation} #1 \end{equation} }

\def\w{\omega}

\def\d{\partial}

\def\sl2r{SL(2,\mathbb{R})}

\def\ce{\varepsilon}

\newcommand{\lsim}{\mathrel{\hbox{\rlap{\lower.55ex \hbox{$\sim$}} \kern-.3em \raise.4ex \hbox{$<$}}}}
\newcommand{\gsim}{\mathrel{\hbox{\rlap{\lower.55ex \hbox{$\sim$}} \kern-.3em \raise.4ex \hbox{$>$}}}}

 \newcommand{\be}{\begin{equation}}
\newcommand{\ee}{\end{equation}}

\usepackage{bm}

\newcommand{\lb}{\left}
\newcommand{\rb}{\right}

\newcommand{\mc}{\mathcal}

\newcommand{\ms}{\mathscr}

\newcommand{\bb}{\mathbb}

\newcommand{\df}[1]{\boldsymbol{#1}}

\newcommand{\defn}{\mathrel{\mathop:}=} 

\makeatletter
\newcommand{\strong}[1]{\@strong{#1}}
\newcommand{\@@strong}[1]{\textbf{\let\@strong\@@@strong#1}}
\newcommand{\@@@strong}[1]{\textnormal{\let\@strong\@@strong#1}}
\let\@strong\@@strong
\makeatother

\begin{document}

\title{\begin{flushright}\vspace{-1in}
       \mbox{\normalsize  EFI-16-17}
       \end{flushright}
       \vskip 20pt
Physical stress, mass, and energy for non-relativistic matter}

\date{\today}

\author{
	Michael Geracie\thanks{\href{mailto:mgeracie@uchicago.edu}         {mgeracie@ucdavis.edu}} \\ 
         {\em  \it  Center for Quantum Mathematics and Physics (QMAP)}\\
   {\it   Department of Physics, University of California, Davis, CA 95616 USA} 
   \and
	Kartik Prabhu\thanks{\href{mailto:kartikprabhu@cornell.edu}         {kartikprabhu@cornell.edu}} \\               {\em \it Cornell Laboratory for Accelerator-based Sciences and Education (CLASSE)}\\
   {\it   Cornell University, Ithaca, NY 14853 USA}
    \and 
	Matthew M. Roberts\thanks{\href{mailto:matthewroberts@uchicago.edu}
     {matthewroberts@uchicago.edu}} \\
 {\em  \it  Kadanoff Center for Theoretical Physics}\\
   {\it   University of Chicago, Chicago, IL 60637 USA}
}

\maketitle

\begin{abstract}
For theories of relativistic matter fields there exist two possible definitions of the stress-energy tensor, one defined by a variation of the action with the coframes at fixed connection, and the other at fixed torsion. These two stress-energy tensors do not necessarily coincide and it is the latter that corresponds to the Cauchy stress measured in the lab. In this note we discuss the corresponding issue for non-relativistic matter theories. We point out that while the physical non-relativistic stress, momentum, and mass currents are defined by a variation of the action at fixed torsion, the energy current does not admit such a description and is naturally defined at fixed connection. Any attempt to define an energy current at fixed torsion results in an ambiguity which cannot be resolved from the background spacetime data or conservation laws. We also provide computations of these quantities for some simple non-relativistic actions.

\end{abstract}

\newpage
\tableofcontents


\section{Introduction}\label{sec:intro}

In relativistic theories with spinful matter there are two possible definitions of the stress-energy tensor \cite{belinfante1940current,Hehl:1976vr,Hehl:2014eja}. One can vary the matter action considering the coframes \(\df e^A \equiv e^A_{\mu}dx^\mu\) and the Lorentz spin connection \(\df \omega^A{}_B \equiv \omega_\mu{}^A{}_B dx^\mu\) as the independent geometric variables
\begin{align}\label{eq:var-S-rel}
	\delta \mc S = \int d^{d+1}x | e| \left( - \tilde T^\mu{}_A \delta e^A_\mu + s^{\mu AB} \delta \omega_{\mu AB}\right)
\end{align}
to define the stress-energy tensor \(\tilde T^\mu{}_A\) and the spin current \(s^{\mu AB}\). Alternatively, since there exists a unique torsion-free metric compatible connection --- the Levi-Civita connection \(\df\omega_{\rm (LC)}{}^A{}_B\) --- which is determined completely by the coframes, one can consider the coframes and the contorsion \(\df C^A{}_B \equiv \df\omega^A{}_B - \df\omega_{\rm (LC)}{}^A{}_B\) as independent
\begin{align}\label{eq:var-S-imp-rel}
	\delta \mc S = \int d^{d+1}x | e| \left( - T^\mu{}_A \delta e^A{}_\mu + s^{\mu AB} \delta C_{\mu AB}\right)
\end{align}
to get
\be\label{eq:tau-imp-rel}\begin{split}
	T^{\mu \nu} & = \tilde T^{\mu \nu} - 2 ( \nabla_\lambda - T^\rho{}_{\rho \lambda} ) S^{\nu \mu \lambda} - T^\mu{}_{\lambda \rho} S^{\nu \lambda \rho}, \\
	\text{where} \qquad \qquad  	S^{\mu\nu\lambda} & = \tfrac{1}{2}\lb( s^{\mu\nu\lambda} - s^{\nu\lambda\mu} - s^{\lambda\mu\nu} \rb) ,
\end{split}\ee
and we have used the Lorentzian coframes and frames to convert the internal frame indices to spacetime indices. In the relativistic case, \eqref{eq:var-S-imp-rel} is equivalent to considering the coframes and the torsion \(\df T^A \equiv \half T^A{}_{\mu\nu} dx^\mu\wedge dx^\nu\) as the independent variables
\begin{align}\label{eq:var-S-imp-T-rel}
	\delta \mc S = \int d^{d+1}x | e| \left( - T^\mu{}_A \delta e^A_\mu + S_A{}^{\mu \nu} \delta T^A{}_{\mu \nu}\right) .
\end{align} 

Note, that the ``new" spin current \(S_A{}^{\mu\nu} = \eta_{AB} e^B_\lambda S^{\lambda\mu\nu}\) is algebraically related (and thus, equivalent) to \(s^{\mu AB}\). But even on torsionless background spacetimes, the ``new" stress-energy tensor \(T^{\mu\nu}\) gets additional contributions from the derivatives of the spin current and is thus, not equivalent to \(\tilde T^{\mu\nu}\) when the matter fields carry spin.

While both \(\tilde T^{\mu\nu}\) and \(T^{\mu\nu}\) are covariant tensors, the latter is the relevant one for most physical problems. The Noether identity corresponding to local Lorentz transformations guarantees that $T^{\mu\nu}$ (but not necessarily $\tilde T^{\mu\nu}$) is symmetric when the background spacetime is torsionless. Thus, the spatial components of \(T^{\mu\nu}\) give rise to the symmetric Cauchy stress tensor which is the relevant physical quantity when considering shearing or straining the system. More directly, stresses in lattice systems are induced by spatial deformations of the system without introducing dislocations i.e.~varying the spatial geometry at fixed torsion.\footnote{See \cite{PhysRevB.92.165131} and references therein for discussions on computing stress response from a lattice theory.} Further, it is known that the symmetric tensor $T^{\mu \nu}$ is the Hilbert stress-energy that couples to gravity through the Einstein equation (see \cite{belinfante1940current,Hehl:1976vr}). \\

The main goal of this paper is to investigate a similar issue that arises for non-relativistic Galilean invariant matter fields with spin and highlight some subtleties not present in the relativistic case.\footnote{Similar results were obtained, using different methods, by \cite{Festuccia:2016awg}.} As many non-relativistic systems are constructed out of particles with spin, this is a crucial step in describing their physical properties in a covariant manner. We summarize the main arguments and results in the following. We work with the covariant construction of non-relativistic spacetimes following the formulation introduced in \cite{GPR-fluids,GPR_geometry}, called Bargmann spacetimes.\footnote{A more thorough list of references for Newton-Cartan geometry and its applications is provided in section~\ref{sec:Bargmann}.} For matter fields on a Bargmann spacetime, the covariant non-relativistic stress-energy is a tensor of the form
\begin{align}\label{eq:tau-decomp}
	 \tau^A{}_I =
	\begin{pmatrix}
		 \varepsilon^0 & -  p_b & -  \rho^0 \\
		 \varepsilon^a & -  T^a{}_b & -  \rho^a
	\end{pmatrix} ,
\end{align}
containing the energy density $ \ce^0$ and current $\varepsilon^a$, the stress tensor $ T^a{}_b$, the momentum $ p_a$, and the mass density $\rho^0$ and current $ \rho^a$. We first define a stress-energy tensor by varying the coframes and Galilean connection as independent variables (i.e. through the analogue of \eqref{eq:var-S-rel})
\eq{
\delta \mc S = \int  d^{d+1} x |e|( - \tilde \tau^\mu{}_I \delta e^I{}_\mu + s^{\mu AB} \delta \w_{\mu AB}) .
}
Here $\df e^I$ contains not just the spacetime coframes $\df e^A$ but also the Newtonian potential $\df a$. We show, for spinful matter fields on torsionless background spacetimes, using the decomposition for \(\tilde \tau^\mu{}_I\) according to \eqref{eq:tau-decomp}, that
	\begin{enumerate*}[label=(\arabic*), ref=(\arabic*)]
	\item the stress tensor \(\tilde T^{ab}\) is not guaranteed to be symmetric  i.e. it does not correspond to the Cauchy stress tensor, and
	\item the momentum need not coincide with the mass current.
	\end{enumerate*}

In contrast to the relativistic case, in general torsionful Bargmann spacetimes one does not have a natural unique reference Galilean connection (unlike the Levi-Civita connection in the relativistic case) and so the Cauchy stress-energy must be computed by varying the coframes and torsion as independent variables (similar to \eqref{eq:var-S-imp-T-rel})
\eq{
\delta \mc S = \int  d^{d+1} x  |e|( \tau^\mu{}_I \delta e^I{}_\mu + S_I{}^{\mu\nu} \delta T^I{}_{\mu\nu})
}
However, due the non-relativistic nature of the spacetime, the variations of the coframes and torsion are not independent but have to satisfy a covariant constraint (see \eqref{eq:temp_torsion}). This constrained variation leads to the following novel feature in non-relativistic theories: The Cauchy stress, momentum, and mass current can be collected into a covariant Cauchy stress-mass tensor,\footnote{A more precise, but unwieldy name, would be the stress-mass-momentum tensor, but as we will show a Noether identity equates the momentum with the mass current for the physical Cauchy stress-mass tensor.}
\eq{	T^{AB}  =
	\begin{pmatrix}
		\rho^0 &  p^b \\
		 \rho^a &  T^{ab}
	\end{pmatrix},
	}
which is unambigously defined in complete analogy with the relativistic case (see \eqref{eq:tau-imp-rel})
\be\begin{split}
	T^{\mu \nu} & = \tilde T^{\mu \nu} - 2(\nabla_\rho - T^\lambda{}_{\lambda\rho}) S^{\rho \mu\nu} - {T^\mu}_{\rho\lambda} S^{\nu \rho\lambda}.
\end{split}\ee
 However, the energy current \(\varepsilon^\mu\) part of \(\tau^\mu{}_I\) is \emph{always} ambiguous up to a choice of covariant antisymmetric tensor (see \eqref{eq:energy-amb}). We argue that this ambiguity is unphysical as the component $\tilde\varepsilon^\mu$ of $\tilde\tau^\mu{}_I$  corresponds to the true internal kinetic energy current\footnote{The kinetic energy current can not be defined in a frame independent way, and in a given local Galilean frame the physical kinetic energy current is $\tilde \tau^\mu{}_0 = \tilde \ce^\mu$. } of a non-relativistic system.

We then show that the Noether identity for local Galilean transformations is
\eq{
T^{[\mu \nu]} = T^{[\mu}{}_{\lambda \rho} S^{\nu ] \lambda \rho }
}
which guarantees that the Cauchy stress tensor \(T^{ab}\) is symmetric when the torsion vanishes $T^\lambda{}_{\mu \nu} = 0$, and that the Cauchy momentum coincides with the Cauchy mass current. Further, the Noether identities for diffeomorphisms give conservation law
\eq{	- e^I_\mu ( D_\nu - {T^\lambda}_{\lambda \nu}) {\tilde \tau^\nu}{}_I  = F_{\mu \nu} j^\nu + R_{AB\mu \nu} s^{\nu AB} - {T^I}_{\mu \nu} {\tilde \tau^\nu}{}_I }
 which contains the work-energy equation (see \eqref{eq:energy_conservation}) as well as a conservation law for the physical stress-mass tensor
\begin{equation}
	(\nabla_\nu - T^\lambda{}_{\lambda \nu}) T^{\nu \mu} = F^\mu{}_\nu j^\nu  + \hat{  \Xi}_A {}^\mu{}_{\nu \lambda} S^{A \nu \lambda} - T^{I \mu}{}_\nu \tilde \tau^\nu{}_I, 
\end{equation}
where $\hat\Xi$ and $T^I$ are proportional to torsion (defined in \eqref{extendedFirstStructure} and \eqref{symIdentity}). Though the torsionful terms are essential for studying energy response \cite{Luttinger:1964zz,Gromov:2014vla} and for applications in non-relativistic fluid dynamics \cite{GPR-fluids}, they do of course vanish in the real world.\footnote{Systems with $d\df n \neq 0$ are necessary to study systems with thermal gradients, for instance when considering Euclidean statistical path integrals with inhomogeneous temperature.} In this case these identities take the simpler form
\begin{align}
	T^{[\mu \nu]} &= 0, \qquad \qquad
	\nabla_\nu T^{\mu \nu} = F^\mu{}_\nu j^\nu.
\end{align}

The remainder of the paper details the above results and is organized as follows. We begin in section \ref{sec:Bargmann} with a summary of Bargmann spacetimes and the relevant geometric data. Section \ref{sec:non-rel} gives explicit formulae for the Cauchy stress, momentum, and mass current in terms of $\tilde \tau^\mu{}_I$ and the spin current $s^{\mu AB}$, and demonstrating the problems inherent in attempting to define  a ``Cauchy energy current''. We give the Noether identities for the Cauchy stress-mass tensor in section \ref{sec:Ward}. In section \ref{sec:ex_matter} we provide examples of Cauchy stress and mass tensors for non-relativistic field theories. Appendix \ref{app:Riemann} collects the symmetry properties of the non-relativistic Reimann tensor in the presence of torsion, which we use to simplify some of the formulae in the main body of the paper.

\section{Bargmann spacetimes}\label{sec:Bargmann}

Newton-Cartan geometry was originally developed by Cartan to describe Newtonian gravity within a geometric framework similar to that of General Relativity \cite{Cartan:1923zea,Cartan:1924yea} (see also \cite{Mal-book, MTW-book}). Recently, it has been used in the condensed matter literature as the natural setting for Galilean invariant physics, with applications that include cold atoms \cite{Son:2005rv}, non-relativistic fluids \cite{GPR-fluids,Jensen:2014ama,Carter:1993aq,Mitra:2015twa}, the quantum Hall effect \cite{Hoyos:2011ez,Son:2013,Golkar:2013gqa,Geracie:2014nka,Geracie:2016inm}, as well as non-relativistic holographic systems \cite{Christensen:2013lma,Christensen:2013rfa,Hartong:2014oma,Hartong:2014pma,Hartong:2015zia}. It is well recognized in the literature that it is necessary to couple these systems to torsionful geometries to define the full suite of currents available in a non-relativistic system and to study their linear response \cite{Christensen:2013rfa,Christensen:2013lma,Hartong:2014pma,Bergshoeff:2014uea,Gromov:2014vla}. Hence in this section and the next, all formulae will be written for the most general case of unconstrained torsion.

A manifestly Galilean covariant definition of torsionful Newton-Cartan geometries was given in \cite{GPR_geometry} (related constructions can be found in \cite{Bergshoeff:2014uea,Hartong:2015wxa,Banerjee:2014pya,Banerjee:2014nja,Banerjee:2015tga,Banerjee:2015rca,Brauner:2014jaa}). These geometries are called Bargmann geometries and this section is dedicated to a brief review of their features. In section \ref{sec:GalReps} we introduce the necessary background, formally define a Bargmann geometry, and collect the identities that will be used repeatedly throughout this note. Section \ref{sec:Physics} then recaps the physics of Bargmann geometries.

\subsection{The Galilean group and its representations}\label{sec:GalReps}

The Galilean group $Gal(d)$, is the set of matrices of the form
\begin{align}
	\Lambda^A{}_B = 
	\begin{pmatrix}
		1 & 0 \\
		- k^a & R^a{}_b 
	\end{pmatrix},
\end{align}
where \(R^a{}_b\) are spatial rotation matrices in \(SO(d)\) and \(k^a\) parametrize Galilean boosts.
Our conventions are that capital Latin indices $A,B,\dots$ transform in the vector representations of $Gal(d)$, while lower case Latin indices $a,b,\dots$ transform under the $SO(d)$ subgroup. The Galilean group preserves the invariant tensors
\begin{align}
	n_A = 
	\begin{pmatrix}
		1 & 0
	\end{pmatrix},
	&&h^{AB} =
	\begin{pmatrix}
		0 & 0 \\
		0 & \delta^{ab}
	\end{pmatrix} ,
	&& \epsilon_{A_0\ldots A_d}.
\end{align}
Here $n_A$ is called the internal {\it clock form}, $h^{AB}$ the internal {\it spatial metric}, and \(\epsilon_{A_0\ldots A_d}\) is the totally antisymmetric symbol with \(\epsilon_{01\ldots d} = 1\). Note that \(h^{AB}\) is degenerate and satisfies
\begin{align}
	h^{AB} n_B = 0.
\end{align}

There is another $(d+2)$-dimensional representation of $Gal(d)$ given by
\be\label{extendedRep}
	\Lambda^I{}_J = 
	\begin{pmatrix}
		1 & 0 & 0 \\
		- k^a & R^a{}_b & 0 \\
		- \frac 1 2 k^2 & k_c R^c{}_b & 1 
	\end{pmatrix}.
\ee
This representation will prove useful in what follows and we call it the \emph{extended representation}. It preserves an extended version of the clock form \(n_I\) as well as a $(d+2)$-dimensional internal metric of Lorentzian signature which we shall use to raise and lower extended indices
\begin{align}
	n_I = 
	\begin{pmatrix}
		1 & 0 & 0 
	\end{pmatrix},
	&&g^{IJ} =
	\begin{pmatrix}
		0 & 0 & 1 \\
		0 & \delta^{ab} & 0 \\
		1 & 0 & 0
	\end{pmatrix} .
\end{align}
The defining and extended representations also together preserve a mixed invariant
\begin{align}
	\Pi^A{}_I =
	\begin{pmatrix}
		1 & 0 & 0 \\
		0 & \delta^a{}_b & 0
	\end{pmatrix},
	&&\Pi^A{}_I = \Lambda^A{}_B \Pi^B{}_J (\Lambda^{-1})^J{}_I
\end{align}
that may be used to project from the extended to the vector representation, or pull back from the covector to the extended representation. For instance
\begin{align}
	n_A \Pi^A{}_I = n_I,
	&&h^{AB} = \Pi^A{}_I \Pi^B{}_J g^{IJ} .
\end{align}

A Bargmann geometry then consists of an extended-valued coframe \(\df e^I\) and a Galilean spin connection \( \df \omega^A{}_B\) valued in the Lie algebra of $Gal(d)$
\begin{align}\label{eq:e-conn}
	\df e^I = 
	\begin{pmatrix}
		\df n \\
		\df e^a \\
		\df a
	\end{pmatrix},
	&& 	\df \omega^A{}_B = 
	\begin{pmatrix}
		0 & 0 \\
		\df \varpi^a & \df \omega^a{}_b
	\end{pmatrix},
\end{align}
where $\df \omega^{(ab)} = 0$.
These transform under the Galilean group as
\begin{align}
	\df e^I \rightarrow \Lambda^I{}_J \df e^J,
	&& \df \omega^A{}_B \rightarrow \Lambda^A{}_C (\df \omega^C{}_D + \delta^C{}_D d) (\Lambda^{-1})^D{}_B. 
\end{align}
We could alternatively present the spin connection in the extended representation as
\begin{align}
	\df \omega^I{}_J = 
	\begin{pmatrix}
		0 & 0 & 0\\
		\df \varpi^a & \df \omega^a{}_b & 0 \\
		0 & - \df \varpi_b & 0
	\end{pmatrix} .
\end{align}
By virtue of being in the Lie algebra of the Galilean group, the connection satisfies the identities
\begin{gather}
	n_A \df \omega^A{}_B = 0,
	\qquad \qquad \df \omega^{(A}{}_{C} h^{B) C} = 0 , \nonumber \\
	n_I \df \omega^I{}_J = 0,
	\qquad \qquad\df \omega^{(IJ)} = 0 ,
	\qquad \qquad \Pi^A{}_J \df \omega^J{}_I = \df \omega^A{}_B \Pi^B{}_I .
\end{gather}
One can then use the Galilean connection to define a Galilean-covariant exterior derivative \(D\) under which the Galilean invariant tensors considered above are covariantly constant
\begin{align}\label{invariantTensors}
	Dn_A = 0,
	&&D h^{AB} = 0,
	&&D n_I = 0,
	&&D g^{IJ} = 0,
	&&D \Pi^A{}_I = 0.
\end{align}

Given this data we may naturally define an extended torsion tensor
\be\label{extendedFirstStructure}
	\df T^I = D\df e^I ,
\ee
which in components reads
\begin{align}	\label{extendedFirstStructureComponents}
	\begin{pmatrix}
		\df T^0 \\
		\df T^a \\
		\df f
	\end{pmatrix} = 	\begin{pmatrix}
		d\df n \\ d\df e^a + {\df \omega^a}_b \wedge \df e^b + \df\varpi^a \wedge \df n    \\ d\df a - \df \varpi_a \wedge \df e^a
	\end{pmatrix},
\end{align}
and transforms covariantly $\df T^I \rightarrow \Lambda^I{}_J \df T^J$. The curvature is, as usual
\begin{align}
	\df R^A{}_B = d \df \omega^A{}_B + \df \omega^A{}_C \wedge \df \omega^C{}_B .
\end{align}

To conclude this section we collect a few further identities that we will use extensively in what follows. 
First, note that the defining and extended representations of the Galilean connection \eqref{eq:e-conn} both contain precisely the same data as a totally antisymmetric matrix of one-forms with lowered indices 
\begin{align}
\label{eq:connection_AB}
	\df {\hat \omega}_{AB} =
	\begin{pmatrix}
		0 & - \df \varpi_b \\
		\df \varpi_a & \df \omega_{ab} 
	\end{pmatrix},
\end{align}
and indeed, they can both be written as
\begin{align}
	\df \omega^A{}_B = h^{AC} \df {\hat \omega}_{CB},
	&&\df \omega^I{}_J = \Pi^{AI} \Pi^B{}_J \df {\hat \omega}_{AB} .
\end{align}
It will often be easier to write equations in terms of $\df{\hat \omega_{AB}}$ rather than $\df \omega^A{}_B$ or $\df \omega^I{}_J$. Under local Galilean transformations $\Lambda(\Theta) = e^\Theta$, it transforms as
\begin{align}\label{omegaTransf}
	\df {\hat \omega_{AB}} \rightarrow (\Lambda^{-1})^C{}_A(\Lambda^{-1})^D{}_B \df {\hat \omega}_{CD} - d \hat \Theta_{AB},
\end{align}
where $\hat \Theta_{AB}$ is the unique antisymmetric matrix such that $\Theta^A{}_B = h^{AC} \hat \Theta_{CB}$.

\subsection{The physics of Bargmann geometries}\label{sec:Physics}
The extended coframe contains the metric data of a Newton-Cartan geometry in its vector part
\begin{align}
	\df e^A = \Pi^A{}_I \df e^I =
	\begin{pmatrix}
		\df n \\
		\df e^a
	\end{pmatrix},
\end{align}
whose components form a basis for the cotangent space of the Galilean spacetime. We can then form the Galilean invariant tensor fields
\begin{align}
	n_\mu = n_A e^A_\mu,
	&&h^{\mu \nu} = e^\mu_A e^\nu_B h^{AB},
\end{align}
where we have introduced the frame fields $e^\mu_A$ satisfying $e^\mu_A e^A_\nu = \delta^\mu{}_\nu$ and \(e^\mu_B e^A_\mu = \delta^A{}_B\). These are the clock-form and spatial metric found in standard treatments of Newton-Cartan geometry \cite{Cartan:1923zea,Cartan:1924yea} and are used to measure elapsed times and spatial distances respectively. 

A spacetime derivative operator $\nabla$ is then defined in the usual way from the connection one-form
\begin{align}
	D_\mu e^A_\nu \equiv \nabla_\mu e^A_\nu + \omega_\mu{}^A{}_B e^B_\nu = 0 .
\end{align}
These satisfy the Newton-Cartan conditions
\begin{align}\label{NCconditions}
	h^{\mu\nu} n_\nu= 0,	&& \nabla_\mu n_\nu = \nabla_\lambda h^{\mu \nu} = 0 
\end{align}
by virtue of the identities \eqref{invariantTensors}.
The vector component $\df T^A$ of the extended torsion gives the spacetime torsion, and \(\df R^A{}_B\) gives the curvature of the derivative operator \(\nabla\).

One of the key features of non-relativistic geometries is that the derivative operator is not determined entirely by the coframes $\df e^A$ since the equation $\df T^A = D \df e^A$ includes an equation of pure constraint
\begin{align}\label{eq:temp_torsion}
	n_A \df T^A = n_A D \df e^A \implies \df T^0 = d \df n .
\end{align}
The final component of the extended equation $\df T^I = D \df e^I$ fixes the remaining freedom in the Galilean connection in terms of $\df a$ and $\df f$. Henceforth we will assume that the derivative operator \(\nabla\) is the one corresponding to a specified extended torsion \(\df T^I\).

The \(1\)-form $\df a$ is the Newtonian gravitational vector potential and it is through the derivative's dependence on $\df a$ that the geometry encodes Newtonian gravity. To see this, consider the case of a metric flat, torsionless spacetime with \(\df n = dt\) and go to a Galilean frame such that $\df a = - \phi dt$ (one may find from \eqref{extendedRep} that such a frame always exists). Then solving the extended first structure equation \eqref{extendedFirstStructure} yields the Christoffel symbols for \(\nabla\)
\begin{align}
	\Gamma^i{}_{tt} = \partial^i \phi ,
\end{align}
the rest being zero. This guarantees that geodesics feel $\phi$ as a Newtonian potential
\eq{
\xi^\nu \nabla_\nu \xi^\mu = 0 \implies \dot{\xi}^i + \xi^j \d_j \xi^i + \d^i \phi = 0,
}
and this is the manner in which a Newton-Cartan geometry encodes Newtonian gravity (see chapter 12 of \cite{MTW-book} for a textbook discussion). The extended component of the torsion $\df f$ is zero on physical, torsionless spacetimes, but is necessary to discuss torsionful spacetimes in a Galilean covariant way. It acts on matter as an external field strength exerting a Lorentz force on mass current $f^\mu{}_\nu \rho^\nu$ (see (\ref{cauchyEquation})).\\

Finally, a Bargmann spacetime also admits a natural volume element
\begin{align}\label{vol}
	\df \varepsilon = \frac{1}{(d+1)!} \epsilon_{A_0 \cdots A_d } \df e^{A_0} \wedge \cdots \wedge \df e^{A_d},
	&&\text{where} 
	&&\epsilon_{01\cdots d} =1.
\end{align}
which may be used to define integration over spacetime. There is similarly a ``volume element" with raised indices
\begin{align}
	\varepsilon^{\mu_0 \cdots \mu_d} = \epsilon^{A_0 \cdots A_d} e^{\mu_0}_{A_0} \cdots . e^{\mu_d}_{A_d} ,
	&&\text{where} 
	&&\epsilon^{01\cdots d} =1.
\end{align}
However, $\varepsilon^{\mu_0 \cdots \mu_d}$ is not $\varepsilon_{\mu_0 \cdots \mu_d}$ with indices raised by $h^{\mu \nu}$ (which would be zero). In local coordinate components
\begin{align}
	\varepsilon_{0 \cdots d} = | e | ,
	&&\varepsilon^{0 \cdots d} = | e |^{-1},
	&&\text{where}
	&&|e| = \text{det} (e^A_\mu) .
\end{align}

\section{Stress-energy for non-relativistic matter fields}\label{sec:non-rel}

In this section we define the stress-energy tensor for non-relativistic theories and discuss the difference between the Cauchy stress and the stress defined at fixed connection. As originally presented in \cite{GPR-fluids}, and as we shall recap in section \ref{sec:StressEnergy}, the non-relativistic stress-energy tensor $\tau^\mu{}_I$ transforms in the extended representation under internal Galilean transformations. This is due to the fact that in non-relativistic theories, energy and mass are not identified and are independent quantities. In addition to the stress and energy currents, this object also contains information on the flow of momentum and mass.

The other key difference with the relativistic case is the constraint
\begin{align}
	\df T^0 = d \df n .
\end{align}
This leads to several complications, as it does not allow us to define a ``Cauchy stress-energy'' directly: any variation of the metric data that includes the clock form \(\df n\) by necessity cannot be done at fixed torsion. There is however an invariant way to isolate the stress, mass, and momentum parts of the stress-energy tensor, each of which admits improvement to the physical tensors. However, as we shall see in section \ref{sec:StressEnergy}, the full stress-energy cannot be improved and any attempt to do so results in an unresolvable ambiguity. We will then demonstrate a way to define the Cauchy stress-mass tensor in section \ref{sec:cauchy-stress-mass}. In section \ref{sec:Hamiltonian} we  discuss why ``improving'' the energy current is unnecessary, as the energy density defined at fixed connection already corresponds to the Hamiltonian density (less terms coupling the system to external potentials).

\subsection{The stress-energy tensor}\label{sec:StressEnergy}

Consider the action for a non-relativistic matter field \(\psi\) which is a functional of \(\psi\) and the background Bargmann geometry given by the extended coframes \(\df e^I\) and the connection \(\hat{\df \omega_{AB}}\) (see \eqref{eq:connection_AB})
\begin{align}
	\mc S [ \psi ; \df e^I , \hat{ \df \omega}_{AB}]
\end{align}
Following \cite{GPR-fluids}, we define the stress-energy tensor $\tilde \tau^\mu{}_I$ and spin-boost current $s^{\mu AB}$ by the variations
\begin{align}\label{eq:varS-unimp}
	\delta \mc S = \int d^{d+1} x \left( - \tilde \tau^\mu{}_I \delta e^I{}_\mu + s^{\mu AB} \delta \hat \omega_{\mu AB} \right) .
\end{align}
where we have assumed that the theory is on-shell so that $\frac{\delta \mc S}{\delta \psi} = 0$. 

Expanding this definition in components \eqref{eq:e-conn}
\begin{align}
	\delta \mc S = \int d^{d+1} x \left( - \tilde \varepsilon^\mu \delta n_\mu + \tilde T^i{}_a \delta e^a_i + \tilde p_a \delta e^a_t + \tilde \rho^\mu \delta a_\mu + s^{\mu ab} \delta \omega_{\mu ab} + b^{\mu  a} \delta \varpi_{\mu  a} \right) ,
\end{align}
where we have denoted the components of $\tilde \tau^\mu{}_I$ and $s^{\mu AB}$ as
\begin{align}
	\tilde \tau^\mu{}_I =
	\begin{pmatrix}
		\tilde \varepsilon^t & - \tilde p_a & - \tilde \rho^t \\
		\tilde \varepsilon^i & - \tilde T^i{}_a & - \tilde \rho^i 
	\end{pmatrix} ,~
	s^{\mu AB} = 
	\begin{pmatrix}
		0 & -\half b^{\mu b} \\
		\half b^{\mu a}& s^{\mu ab} 
	\end{pmatrix} .
\end{align}
The object $\tilde \tau^\mu{}_I$ then collects the energy current $\tilde \varepsilon^\mu$, the stress $\tilde T^i{}_a$, the momentum $\tilde p^a$, and the mass current $\tilde \rho^\mu$ into a single covariant object that transforms in the extended representation under internal Galilean transformations. We shall call $\tilde \tau^\mu{}_I$ the stress-energy tensor, though as we have seen it contains far more information than the name suggests. Similarly the spin-boost current \(s^{\mu AB}\) collects together the boost current \(b^{\mu a}\) and the spin current \(s^{\mu ab}\).

It is possible to isolate certain components of $\tilde \tau^\mu{}_I$ in a covariant way. For instance, the \emph{stress-mass} tensor may be defined as
\begin{align}
	\tilde T^{AB} = - \tilde \tau^{A}{}_I \Pi^{BI} =
	\begin{pmatrix}
		\tilde \rho^0 & \tilde p^b \\
		\tilde \rho^a & \tilde T^{ab}
	\end{pmatrix}
\end{align}
and contains all of the currents of $\tilde \tau^\mu{}_I$ except the energy current. Similarly, the mass current can be isolated as
\begin{align}\label{rhoEq}
	\tilde \rho^A = \tilde T^{AB} n_B = - \tilde \tau^A{}_I n^I =
	\begin{pmatrix}
		\tilde \rho^0 \\
		\tilde \rho^a
	\end{pmatrix}.
\end{align}

However one cannot isolate the energy current $\tilde \varepsilon^A$ from any of the other currents in $\tilde \tau^A{}_I$ without additional data since it picks up factors of the stress, momentum, and mass currents under a local Galilean boost transformation
\begin{gather}
	\tilde \varepsilon^0 \mapsto \tilde \varepsilon^0 + k_a \tilde p^a + \frac 1 2 k^2 \tilde \rho^0 , \nonumber \\
	\tilde \varepsilon^a \mapsto \tilde \varepsilon^a + \tilde T^{ab} k_a + \frac 1 2 k^2 \tilde \rho^a .\label{energyTransf}
\end{gather}
This is to be expected on physical grounds since the energy current also includes the kinetic energy of the system, which depends explicitly on a notion of rest frame.

However, given a family of observers with velocity $v^\mu$, normalized so that $n_\mu v^\mu = 1$, one may discuss the energy current as measured by these observers as follows. Let $\mathring v^I$ denote the unique null extension of $v^\mu$ to the extended representation. That is
\begin{align}
	v^\mu = \Pi^\mu{}_I \mathring v^I,
	&&\mathring v_I \mathring v^I = 0 ,
	\quad \implies \mathring v^I = 
		\begin{pmatrix} 1 \\ v^a \\ - \frac 1 2 v^2 \end{pmatrix}
\end{align}
Then, the energy current measured by the observer moving with velocity \(v^\mu\) is given by
\begin{align}\label{comovingEnergyCurrent}
	\overset v {\tilde \varepsilon} {}^\mu = \tilde \tau^\mu{}_I \mathring v^I .
\end{align}

While compact, this definition may seem somewhat obtuse. To lend some motivation, we compute the relationship between the energy measured by an observer $v^\mu$ as defined above and the energy as measured in the lab frame, that is, the component $\tilde \varepsilon^0 = \tilde \tau^0{}_{0}$ of the stress-energy tensor. For simplicity we will consider the flat, spinless case, and so drop the tildes
\begin{gather}
		\varepsilon^0 
	=
	v_b \rho^b - \frac 1 2 \rho^0 v^2 + \overset v {\varepsilon }{}^0 .
\end{gather}
In writing this we have also used the upcoming Ward identity (\ref{eq:noether-gal-imp}) for local Galilean transformations, which in particular implies $p^i = \rho^i$. This looks more familiar if we consider the case of an observer comoving with the mass current. In this case we have $\rho^a = \rho^0 v^a$ and the above simplifies to 
\begin{gather}\label{nonCovCur}
		\varepsilon^0 
	=
		\frac 1 2 \rho^0 v^2 + \overset v \varepsilon {}^0 .
\end{gather}
 The lab energy density $\varepsilon^0$ then includes both the energy density $\overset v \varepsilon {}^0$ as measured by observers in the rest frame defined by $v^\mu$, plus the kinetic energy $\frac 1 2 \rho^0 v^2$ that the comoving observers do not measure.

As we will show in \eqref{eq:noether-gal-unimp}, for spinful matter fields on torsionless spacetimes, the Ward identity for local Galilean transformations (on torsionless spacetimes) is
\be
	\tilde T^{[\mu \nu]} = - \nabla_\lambda s^{\lambda \mu \nu} .
\ee
For spinful matter, we see that the stress tensor \(\tilde T^{ij}\) need not be symmetric and the mass current \(\tilde\rho^i\) may not coincide with the momentum \(\tilde p^i\). To get the appropriately symmetric stress-energy we can proceed in analogy to the relativistic case discussed in section \ref{sec:intro}. However, on Bargmann spacetimes with torsion there is no natural reference connection analogous to the Levi-Civita connection in the relativistic case; hence, there is no analog of the variation section \ref{eq:var-S-imp-rel}.

Thus, to get the symmetric (i.e. Cauchy) stress tensor for non-relativistic fields we should vary the action considering the extended coframe and extended torsion as the independent geometric variables (the analogue of \eqref{eq:var-S-imp-T-rel}). However, due to the identity \eqref{extendedFirstStructure} (in particular \eqref{eq:temp_torsion}), this amounts to doing a constrained variation since the variations must satisfy 
\be\label{eq:var-constraint}
	n_I \delta \df T^I = d\delta \df n .
\ee

To carry out this constrained variation of the action we first note that, from \eqref{eq:var-constraint}, the variation of the torsion holding the coframes \(\df e^I\) fixed satisfies \(n_I \lb. \delta \df T^{I} \rb\vert_{\df e^I} = 0\). Thus, there is a unique two-form \(\delta \hat{\df T}_A\) satisfying \(\lb. \delta \df T^{I} \rb\vert_{\df e^I} = \Pi^{AI} \delta \hat{\df T}_A\). From $\delta \df T_I = ( \delta \df f,  \delta \df T_a, d \delta \df n )$ we can get the explicit expression
\be\label{eq:T-hat}
	\delta \hat{\df T}_A =
		\lb. \begin{pmatrix}
			\delta \df f \\ 
			\delta \df T_a
		\end{pmatrix}\rb\vert_{\df e^I} .
\ee

Now we would like to vary the coframes at fixed $\df T^I$. However, fixing the torsion implies the constraint $d \delta \df n =0$. Due to the constrained nature of the variation, this cannot fully fix the stress-energy tensor without additional data. Let us see how this works. We try to write the variation of the action in the form
\be\label{eq:varS-imp}
	\delta\mc S = \int  d^{d+1} x  |e|\left( -{\tau^\mu}_I \lb. \delta e^I_\mu \rb\vert_{\df T} + S^{A\mu\nu} \delta \hat T_A{}_{\mu\nu} \right) .
\ee
This is what we would like to consider the physical (or Cauchy) stress-energy tensor. 

As before, we can isolate the stress-mass components
\begin{align}
	T^{A B} = - \tau^{A}{}_I \Pi^{B I} = 
	\begin{pmatrix}
		\rho^0 & p^b \\
		\rho^a & T^{ab}
	\end{pmatrix} .
\end{align}
As we shall show (see \eqref{eq:noether-gal-imp}), this is symmetric on torsionless spacetimes, so that the stress tensor \(T^{ij}\) is symmetric and the momentum \(p^i\) and mass currents \(\rho^i\) coincide. Thus, we call the stress-mass tensor \(T^{AB}\) obtained from the variation \eqref{eq:varS-imp} the \emph{Cauchy stress-mass tensor}.\\

The ambiguity in $\tau^\mu{}_I$ that results from the constraint \eqref{eq:var-constraint} can be classified as follows. Consider an arbitrary antisymmetric tensor \(H^{\mu\nu} = H^{[\mu\nu]}\) constructed from the field \(\psi\), the geometric data \(\df e^I\) and \(\df\omega^I{}_J\), and their derivatives. Then, to \eqref{eq:varS-imp} we can always add the following term, which does not affect the variation of the action since we must vary the torsion keeping the coframes fixed
\begin{align}
	-\int  d^{d+1} x  |e| H^{\mu\nu} n_I \delta T^I_{\mu\nu}  &= - \int  d^{d+1} x |e| H^{\mu\nu} (d\delta n)_{\mu\nu} \nonumber \\
		&= 2 \int  d^{d+1} x  |e| H^{\mu\nu} \left(\nabla_\nu\delta n_\mu + \tfrac{1}{2}{T^\lambda}_{\nu\mu}\delta n_{\lambda} \right) .
\end{align}
Integrating by parts and ignoring the boundary term, this becomes
\be
	- \int d^{d+1} x |e| \left( 2 (\nabla_\nu - T^\lambda{}_{\lambda\nu}) H^{\mu\nu} + {T^\mu}_{\nu\lambda}H^{\nu\lambda} \right) \delta n_\mu .
\ee
Thus the \(\tau^\mu{}_I\) is ambiguous up to the redefinition
\be\label{eq:tau-amb}
	{\tau^\mu}_I \rightarrow {\tau^\mu}_I + \left( 2 (\nabla_\nu - T^\lambda{}_{\lambda\nu}) H^{\mu\nu} + {T^\mu}_{\nu\lambda}H^{\nu\lambda} \right) n_I .
\ee
Since the ambiguity is proportional to $n_I$ this only affects the energy current components of $\tau^\mu{}_I$
\be \label{eq:energy-amb}
	\varepsilon^\mu \rightarrow \varepsilon^\mu + 2 (\nabla_\nu - T^\lambda{}_{\lambda\nu}) H^{\mu\nu} + {T^\mu}_{\nu\lambda}H^{\nu\lambda} ,
\ee
while the Cauchy stress-mass tensor \(T^{\mu\nu}\) is unambiguously defined. One might hope that conservation laws might resolve this ambiguity, but from \eqref{eq:energy_conservation} and the fact that the ambiguity is the divergence of an antisymmetric tensor we find it does not.

Thus, while the physical stress, mass, and momentum are contained within the Cauchy stress-mass tensor $T^{\mu \nu}$, the physical energy current is the ``unimproved'' object $\tilde \varepsilon^\mu$. It transforms under internal Galilean transformations according to \eqref{energyTransf}. Fortunately we will derive a relationship between the components of $\tilde T^{AB}$ and $T^{AB}$ and therefore can rewrite \eqref{energyTransf} purely in terms of physical quantities.

%
\subsection{The Cauchy stress-mass tensor}\label{sec:cauchy-stress-mass}

Let us now consider the stress-mass tensor, whose story is straightforward. The physical currents are to be defined at fixed torsion, so we will require that the variation of $\df e^I$ does not involve the clock form, which is fine as we are neglecting energy currents. This in turn implies that the variation may be written as the pullback of some form $\delta \df {\hat e}_A$ to the extended representation 
\begin{align}
	n_I \delta \df e^I = 0 \implies \delta \df e^I = \Pi^{AI} \delta \df \hat{\df e}_A .
\end{align}
Similarly we also have
\begin{align}
	n_I \delta \df T^I = 0 \implies \delta \df T^I  = \Pi^{A I} \delta \df{\hat T}_{A}
\end{align} 
for some $(\delta \df {\hat T}_{A})_{ \mu \nu}$. We then define the Cauchy stress-mass $T^{\mu \nu}$ and spin-boost current $S^{A\mu \nu}$ by the variation
\begin{align}\label{improvedCurrentsDef}
	\delta \mc S = \int  d^{d+1} x | e | \left( T^{\mu A} (\delta {\hat e}_A)_{ \mu} + S^{A \mu \nu}  (\delta \hat T_A)_{ \mu \nu}\right) .
\end{align}

To perform the translation between \eqref{eq:varS-unimp} and \eqref{improvedCurrentsDef} we shall need the variation of the spin connection $\delta \hat \omega_{\mu AB}$ in terms of $(\delta\hat e_A)_\mu$ and $(\delta {\hat T}_A)_{ \mu \nu}$. We can retrieve this from the first structure equation $\df T^I = D \df e^I$, which gives
\begin{align}
	\delta\df  T^I = D \delta \df e^I + \delta \df \omega^I{}_J \wedge \df e^J .
\end{align}
Using $\delta \df e^I = \Pi^{AI} \delta \df {\hat e}_A$, $\delta \df T^I = \Pi^{AI} \delta \df {\hat T}_A $, and $\delta \df \omega_{IJ} = \Pi^A{}_I \Pi^B{}_J \delta\df  {\hat \omega}_{AB}$, this reads
\begin{align}
	\delta\df  {\hat \omega}_{AB} \wedge \df e^B = \delta \df{\hat T}_A  - D \delta\df  {\hat e}_A ,
\end{align}
which after some algebraic rearrangement gives
\begin{align}
	\delta \hat \omega_{CAB} &= \delta \hat \omega_{[CA]B} + \delta \hat \omega_{[BC]A}-\delta \hat \omega_{[AB]C} \nonumber \\
		&= \frac 1 2 \bigg(  ( D \delta {\hat e}_A )_{BC} + (D \delta { \hat e}_B)_{CA} - ( D \delta {\hat e}_C )_{AB} \nonumber \\
		& \qquad \qquad \qquad - ( \delta {\hat T}_A)_{BC} - (\delta {\hat T}_B)_{CA} + (\delta {\hat T}_C)_{AB} \bigg) .
\end{align}

Finally, using $\delta \df e^I = \Pi^{AI} \delta \df e_A$ and the definition $\tilde T^{AB} = - \Pi^{AI} \tilde \tau^A{}_I$, we find that
\begin{align}
	\delta \mc S &= \int d^{d+1} x | e | \left( \tilde T^{\mu A} \delta e_{A \mu} +  s^{\mu AB} \delta \hat \omega_{\mu AB}\right) \nonumber \\
	&= \int d^{d+1} x | e | \left( T^{\mu A} (\delta {\hat e}_A)_{ \mu} + S^{A \mu \nu} (\delta {\hat T}_A)_{ \mu \nu}\right) ,
\end{align}
along with the relations
\begin{align}\label{stressMomentumImprovement}
	S^{\lambda \mu \nu} & =  \frac 1 2 \left(  s^{\lambda \mu \nu}  -  s^{\mu \nu \lambda} -  s^{\nu \lambda \mu} \right), \nonumber \\
	T^{\mu \nu} & = \tilde T^{\mu \nu} - 2 ( \nabla_\lambda - T^\rho{}_{\rho \lambda} ) S^{\nu \mu \lambda} - T^\mu{}_{\lambda \rho} S^{\nu \lambda \rho} .
\end{align}
In particular, this gives a physical mass current $\rho^\mu = T^{\mu \nu} n_\nu$
\begin{align}\label{rhoImprovement}
	\rho^\mu = \tilde \rho^\mu -2 ( \nabla_\lambda - T^\rho{}_{\rho \lambda} ) S^{\nu \mu \lambda}n_\nu - T^\mu{}_{\lambda \rho} S^{\nu \lambda \rho} n_\nu .
\end{align}
This is the physical mass current for spinful matter, it is $\rho^\mu$ rather than $\tilde \rho^\mu$ that flows in response to a gravitational perturbation at fixed torsion.


\subsection{Energy currents and the Hamiltonian}\label{sec:Hamiltonian}

Let us directly confirm that the component $\tilde \tau^t{}_0$ of the stress-energy \(\tilde \tau^\mu{}_I\) is truly the (kinetic) energy density of a simple spinful theory, the spinful Schr\"odinger equation, whose action on the background $n_\mu = ( n_t, ~ 0),~ h^{\mu \nu} = \begin{pmatrix} 0 & 0 \\ 0 & \delta^{ij} \end{pmatrix}$ is
\begin{gather}
	\mc S = \int d^{d+1} x \sqrt{h} n_t \left( \frac{1}{n_t} \frac i 2 \psi^\dagger \overset{\leftrightarrow} D{}_t \psi - \frac{\delta^{ij}}{2m} D_i \psi^\dagger D_j \psi \right),
	\nonumber \\
	\text{where}
	\qquad \qquad
	D_\mu \psi = \left( \partial_\mu - i q A_\mu - i m a_\mu - \frac{i}{2} J^{ab} \omega_{ab} \right) \psi
\end{gather}
and $J^{ab}$ are the spin representation matrices. One then finds
\begin{align}
	\tilde\varepsilon^t = - \frac{\delta S}{\delta n_t} = \frac{\delta^{ij}}{2m} D_i \psi^\dagger D_j \psi  ,
\end{align}
whereas the Hamiltonian density for this system is
\begin{align}\label{schrodH}
	\mc H = \frac{\delta^{ij}}{2m} D_i \psi^\dagger D_j \psi - q A_t \psi^\dagger \psi - m a_t \psi^\dagger \psi - \frac 1 2 \omega_{t ab} \psi^\dagger J^{ab} \psi .
\end{align}
We see that $\tilde\varepsilon^t = \tilde \tau^t{}_0$ is the Hamiltonian density minus coupling to external potentials and so corresponds to the {\it internal} kinetic and interaction energy of a system.

While we have motivated this in the specific case of the Schr\"odinger theory, a similar analysis shows that it is the energy density \(\tilde \varepsilon^t\) that enters the Hamiltonian $\mathcal H$ and provides a generalization of (\ref{schrodH}) for arbitrary theories.

\section{Ward identities}\label{sec:Ward}

The stress-energy tensor \(\tilde \tau^\mu{}_I\) and the spin-boost current \(s^{\mu AB}\) satisfy certain Ward identities by virtue of the action begin invariant under diffeomorphisms, local \(U(1)_{\mathbb{M}}\) transformations, and local Galilean transformations. These were computed in a manifestly covariant form in section 5 of \cite{GPR-fluids}, following derivations in flat space in \cite{Son:2008ye,Janiszewski:2012nb} and in non-covariant form on curved space in \cite{Geracie:2014nka}. In the spinful case, these Ward identities were for the unimproved currents defined at fixed connection. In this section we present the corresponding identities for the Cauchy stress-mass tensor $T^{\mu \nu}$ and energy current $\tilde \ce^\mu$.

In \cite{GPR-fluids}, we found that invariance of the action under local \(U(1)_{\mathbb{M}}\) transformations and diffeomorphisms gives us the conservation laws
\begin{subequations}\begin{align}
	(\nabla_\mu - T^\nu{}_{\nu \mu}) \tilde \rho^\mu & = 0 \label{massCons}\\
	- e^I_\mu ( D_\nu - {T^\lambda}_{\lambda \nu}) {\tilde \tau^\nu}{}_I & = F_{\mu \nu} j^\nu + R_{AB\mu \nu} s^{\nu AB} - {T^I}_{\mu \nu} {\tilde \tau^\nu}{}_I \label{diffWard} ,
\end{align}\end{subequations}
where we have also include an external electromagnetic field $F_{\mu \nu}$ coupling to the charge-current $j^\mu$.
Raising the index on \eqref{diffWard} with $h^{\mu \nu}$and using the identity
\eq{
e^{I\mu} = \Pi^{\mu I} + (a^\mu - e^\mu_0) n^I}
 we also find for the stress-mass tensor
\begin{align}
	( \nabla_\nu - {T^\lambda}_{\lambda \nu}) {\tilde T^{\nu \mu}} = F^\mu{}_{\nu} j^\nu + R_{AB}{}^\mu{}_\nu {s^{\nu AB}} - {T^{I\mu}}_{\nu} {\tilde \tau^\nu}{}_I.
\end{align}

The equation \eqref{massCons} is simply the conservation of mass on torsionful spacetimes, while \eqref{diffWard} is a covariant version of energy conservation and the continuum version of Newton's second law (also called the Cauchy momentum equation; see \eqref{eq:Cauchy-mom-eqn}). To make this more transparent, restrict to spinless matter on flat, torsionless spacetimes in Cartesian coordinates \((t, x^i)\), in the presence of a Newtonian gravitational potential \(\phi\). In an inertial frame, we have $\df e^I = ( dt, ~ d x^i, ~ - \phi dt )$, and churning through the temporal and spatial components of (\ref{massCons},\ref{diffWard}), we find
\begin{subequations}\label{flatSpaceFluid}\begin{align}
	\dot { \rho}^t + \partial_i  \rho^i & = 0 \\
	\dot { \varepsilon}^t + \partial_i  \varepsilon^i & = E_i j^i -\partial_i \phi~  \rho^i,\\
	\dot {  p}^i + \partial_j  T^{ij} & = E^i j^t + \epsilon^{ijk} j_j B_k -\partial^i\phi~  \rho^t \label{eq:Cauchy-mom-eqn}.
\end{align}\end{subequations}
where \(E_i\) and \(B_i\) are the external electric and magnetic fields respectively.

\subsection{Galilean Ward identity}\label{sec:GalWard}

In this section, we consider the Ward identity that follows from the invariance of the action under local Galilean transformations. This has previously been discussed in a non-coviariant form in \cite{Geracie:2014nka,Jensen:2014aia,GPR-fluids}, and we take to opportunity here to finally state the covariant version, from which we derive the symmetry of the Cauchy stress-mass on torsionless spacetimes. Under infinitesimal Galilean transformations we have
\begin{align}\label{infinitesimalGal}
	\delta \df e^I = \Pi^{AI} \hat \Theta_{AB} \df e^B, 
	&& \delta \hat{\df \omega}_{AB} = - D \hat \Theta_{AB} .
\end{align}
where $\Theta^I{}_J = \Pi^{AI} \Pi^B{}_J \hat \Theta_{AB}$.

Local Galilean invariance of the action then implies
\begin{align}
	0 = \delta \mc S &= \int d^{d+1} x | e | \left( - \tilde \tau^\mu{}_I \Pi^{AI} \hat \Theta_{AB} e^B_\mu - s^{\mu AB} D_\mu \hat \Theta_{AB} \right) \nonumber \\
		&= \int d^{d+1} x | e | \hat \Theta_{AB} \left( \tilde T^{AB} + (D_\mu - T^\nu{}_{\nu \mu} ) s^{\mu AB}\right) ,
\end{align}
from which we find the Ward identity
\be\label{eq:noether-gal-unimp}
	\tilde T^{[\mu \nu]} = - ( \nabla_\lambda - T^\rho{}_{\rho \lambda} ) s^{\lambda \mu \nu} .
\ee
Thus, for spinful matter, even on torsionless backgrounds that preserve local rotational invariance, \(\tilde T^{\mu\nu}\) fails to be symmetric.  \(\tilde T^{ij}\) is then not the Cauchy stress tensor used commonly in physics and engineering applications \cite{irgens2008continuum}. Moreover, the momentum \(\tilde p^i\) need not coincide with the mass current \(\tilde \rho^i\) in the presence of inhomogeneous spinful matter, violating a common constraint assumed in non-relativistic physics \cite{Greiter:1989qb}. 

Both of these conditions do however hold for the Cauchy stress-mass $T^{\mu \nu}$, which by virtue of \eqref{eq:noether-gal-unimp} and \eqref{stressMomentumImprovement} satisfies the Ward identity
\begin{align}\label{eq:noether-gal-imp}
	T^{[\mu \nu]} = T^{[\mu}{}_{\lambda \rho} S^{\nu] \lambda \rho } ,
\end{align}
so that on torsionless backgrounds we have $T^{[\mu \nu ]} =0$ . This guarantees \(p^i = \rho^i\), generalizing the relation $T^{0i} = \frac{m}{e} j^i$ used by Greiter, Witten, and Wilczek \cite{Greiter:1989qb} and subsequent authors \cite{Son:2005ak} to impose Galilean invariance to the case of multi-constituent systems. Note that due to a manifestly covariant formalism this relationship is guaranteed and we do not need to impose it as a functional constraint on the effective action as in \cite{Greiter:1989qb}.

\subsection{Diffeomorphism and $U(1)_{\mathbb{M}}$ Ward identity }

We would now like to state the diffeomorphism Ward identity
\begin{align}
	- e^I_\mu ( D_\nu - {T^\lambda}_{\lambda \nu}) {\tilde \tau^\nu}{}_I & = F_{\mu \nu} j^\nu + R_{AB\mu \nu} s^{\nu AB} - {T^I}_{\mu \nu} {\tilde \tau^\nu}{}_I 
\end{align}
in terms of the physical currents as much as possible. There is unfortunately nothing that can be done about the full equation as it stands since, as we have seen, there is no way to improve the stress-energy tensor as a whole. We can however do so for the Cauchy-momentum equation
\begin{align}\label{cauchyWardUnimp}
	( \nabla_\nu - {T^\lambda}_{\lambda \nu}) {\tilde T^{\nu \mu}} = F^\mu{}_{\nu} j^\nu + R_{AB}{}^\mu{}_\nu {s^{\nu AB}} - {T^{I\mu}}_{\nu} {\tilde \tau^\nu}{}_I
\end{align}
that follows from it and the conservation of $\tilde \rho^\mu$.

Using \eqref{stressMomentumImprovement}, we find this reads
\begin{align}\label{cauchyIntermediate}
	(\nabla_\nu - T^\lambda{}_{\lambda \nu}) T^{\nu \mu} = F^\mu{}_\nu j^\nu  +(  2 \hat R_{ \rho \nu \lambda}{}^\mu - R^\mu{}_{\rho \nu \lambda}  ) s^{\rho \nu \lambda} - T^{I \mu}{}_\nu \tilde \tau^\nu{}_I ,
\end{align}
Here $\hat R_{ABCD}$ is the unique tensor antisymmetric in its first two indices such that $R^A{}_{BCD} = h^{AE}\hat R_{EBCD}$ given by
\begin{align}
	\hat {\df R}_{AB} = d \hat {\df \omega}_{AB} + \hat {\df \omega}_{AC} \wedge \hat {\df \omega} {}^C{}_B .
\end{align}
We have also used the identity $2 R_{[ \mu \nu]} = 3 \nabla_{[\mu} T^\lambda{}_{\lambda \nu]} + T^\lambda{}_{\lambda \rho} T^\rho{}_{\mu \nu}$ to simplify the result.

We now simplify the second term on the right hand side of \eqref{cauchyIntermediate} using the symmetry of the Riemann tensor under exchange of the first and second pairs of indices. This identity is slightly more subtle than the usual relativistic case since we do not have an invertible metric tensor. We first note that
\begin{align}\label{TAdef}
	n_I D \df T^I = d^2 \df n = 0 \implies D \df T^I = \Pi^{AI} \hat{ \df \Xi}_A
\end{align}
for some \(2\)-form $\hat{ \df \Xi}_A$ whose components read
\begin{align}
	\hat{ \df \Xi}_A = 
	\begin{pmatrix}
		d \df f - \df \varpi^b \df \wedge \df T_b ,&
		d \df T_a - \df T_b \wedge \df \omega^b{}_a + \df T^0 \wedge \df \varpi_a
	\end{pmatrix} .
\end{align}
While $\hat{ \df \Xi}_A$ is not covariantly exact, its raised index version is simply
\begin{align}
	D\df T^A = h^{AB} \hat{ \df \Xi}_B ,
\end{align}
Using this, the identity for the symmetry of the Riemann tensor under exchange of the first and second pairs of indices is given by
\begin{gather}\label{symIdentity}
	\hat R_{ABCD} = \hat R_{CDAB} + \frac 1 2 \left( \hat{  \Xi}_{ABCD} +  \hat{  \Xi}_{BCAD} +  \hat{  \Xi}_{CADB} + \hat{  \Xi}_{DABC} \right) ,
\end{gather}
where $\hat{  \Xi}_{ABCD}=(\hat{  \Xi}_A )_{\mu\nu\rho}e^\mu_B e^\nu_C e^\rho_D $. The interested reader can find the proof of (\ref{symIdentity}) in Appendix \ref{app:Riemann}.

Using \eqref{symIdentity} we then find that (\ref{cauchyIntermediate}) simplifies to
\begin{align}\label{cauchyEquation}
	(\nabla_\nu - &T^\lambda{}_{\lambda \nu}) T^{\nu \mu} = F^\mu{}_\nu j^\nu  + \hat{  \Xi}_A{}^\mu{}_{\nu \lambda} s^{A \nu \lambda} - T^{I \mu}{}_\nu \tilde \tau^\nu{}_I ,
\end{align}
which is the covariant generalization of the Cauchy momentum equation to unconstrained Bargmann spacetimes. In particular we see that there are external forces exerted by extended torsion on spin current and stress-energy, in addition to the usual Lorentz force on $j^\mu$. While it may seem awkward to include the unimproved tensor $\tilde \tau^\mu{}_I$ in (\ref{cauchyEquation}), having converted everything else to the physical currents, this is something we must simply accept as we have shown there is no unambiguous way to improve it. We simply observe that the external force exerted by extended torsion couples to the unimproved stress-energy. It is of course possible, to decompose $\tilde \tau^\mu{}_I$ into $\overset v {\tilde \varepsilon}{}^\mu$ and $\tilde T^{\mu \nu}$ and then convert to the Cauchy stress-mass tensor $T^{\mu \nu}$, but at the cost of introducing a preferred frame $v$.

Unlike the Cauchy equation \eqref{cauchyEquation}, the work-energy equation cannot be isolated in a Galilean frame independent manner. The problem is that observed in the discussion following \eqref{rhoEq}: while one can invariantly isolate the stress-mass part of the the stress-energy tensor, there is no observer independent definition of energy. This is to be expected on physical grounds since the energy current also includes the kinetic energy of the system, which must be defined with respect to some notion of rest. However, given a family of observers with velocity $v^\mu$, normalized so that $n_\mu v^\mu = 1$, one may define the energy current as measured by these observers to be \eqref{comovingEnergyCurrent}, which we reproduce here
\begin{align}
	\overset v {\tilde \varepsilon} {}^\mu = \tilde \tau^\mu{}_I \mathring v^I .
\end{align}

Now we saw previously in \eqref{flatSpaceFluid} that the temporal component of the diffeomorphism Ward identity contains the work-energy equation. Given a family of observers, we can obtain the covariant version of this by contracting the Ward identity with some frame $v^\mu$. In doing so, the following identity is useful
\begin{align}\label{extendedContract}
	e^I_\mu v^\mu = \mathring v^I + \mathring v_J e^J_\mu v^\mu n^I.
\end{align}
Using this equation and mass conservation, one finds 
\begin{align}
	\mathring v^I (D_\mu - T^\nu{}_{\nu\mu} ) \tilde\tau^\mu{}_I &= F_{\mu \nu} j^\mu v^\nu + R_{AB \mu \nu} s^{\mu AB} v^\nu + T^I{}_{\mu \nu} v^\mu \tilde \tau^\nu{}_I, \nonumber \\
	\implies \qquad
	( \nabla_\mu - T^\nu{}_{\nu \mu} ) \overset v {\tilde \varepsilon} {}^\mu &= F_{\mu \nu} j^\mu v^\nu + R_{AB \mu \nu} s^{\mu AB} v^\nu + T^I{}_{\mu \nu} v^\mu \tilde \tau^\nu{}_I - \tau^\mu{}_I D_\mu \mathring v^I .
\end{align}

To simplify the final term, we note that $n_I D_\mu \mathring v^I = \mathring v_I D_\mu \mathring v^I = 0$, so there is a $v^\nu$ orthogonal tensor $t_{\mu \nu}$ such that $D_\mu \mathring v^I = \Pi^{\lambda I} t_{\mu \lambda}$. Contracting this equation with $\Pi^\nu{}_I$ we find that $t_{\mu}{}^\nu = \nabla_\mu v^\nu$ and so
\begin{align}\label{extendedShear}
	D_\mu \mathring v^I = \overset v h_{\nu \lambda} \Pi^{\lambda I} \nabla_\mu v^\nu 
\end{align}
where $\overset v h_{\mu \nu}$ is the unique $v^\mu$ orthogonal symmetric tensor such that $h^{\mu \lambda} \overset v h_{\lambda \nu}  = \delta^\mu{}_\nu - v^\mu n_\nu$.
Plugging this in, one finds the work-energy equation for the comoving energy current is
\begin{align}
\label{eq:energy_conservation}
( \nabla_\mu - T^\nu{}_{\nu \mu} ) \overset v {\tilde\varepsilon} {}^\mu &= F_{\mu \nu} j^\mu v^\nu + R_{AB \mu \nu} s^{\mu AB} v^\nu + T^I{}_{\mu \nu} v^\mu \tilde \tau^\nu{}_I - \tilde T^{\mu \nu} \overset v h_{\nu \lambda} \nabla_\mu  v^\lambda.
\end{align}

\section{Examples}\label{sec:ex_matter}

Finally, let us turn to a few examples. In this section we collect computations for the stress-energy, spin current, and Cauchy stress-mass for various non-relativistic field theories. The principle aim of this discussion will be to derive covariant formulae for these objects and to demonstrate how to carry out the computation maintaining manifest covariance throughout. 

We begin with the spinful Schr\"{o}dinger field in section \ref{sec:sch}. The formulae \eqref{eq:cauchy_schrodinger}  we derive, in their flat space component form \eqref{eq:schrodinger_component}, should for the most part be familiar, but also include spin contributions to the Cauchy stress tensor and mass current which to our knowledge are not present in the literature. In section \ref{sec:Dirac} we consider the non-relativistic Dirac theory which is a Galilean invariant theory for matter charged under both boosts and spatial rotations and is first order in both time and spatial derivatives.  We conclude with the Wen-Zee term which arises in the effective actions for describing quantum Hall states.


\subsection{Spinful Schr\"odinger field}\label{sec:sch}

We begin by considering a massive spinful field \(\psi\) with dynamics given by the Schr\"odinger action. To write a Schr\"odinger action for \(\psi\) it will be essential that the representation of the Galilean algebra on \(\psi\) be unitrary, which restricts \(\psi\) to be invariant under Galilean boosts.\footnote{In the standard treatment, one specifies that $\psi$ transforms projectively under boost transformations, picking up a phase factor $e^{i \frac 1 2 m k^2 t - i m k_i x^i}$. This method of imposing Galilean invariance cannot be used in a curved spacetime as there do not exist global inertal coordinates. As detailed in section 1.2 of \cite{GeracieThesis} and 2.2 of \cite{Geracie:2016inm}, one may view the phase factor as an attempt to absorb the boost transformation of $a_\mu$ into a $U(1)_\mathbb{M}$ transformation for \(\psi\). As such, the phase factor does not appear in our treatment and a Schr\"odinger field is trivial under local Galilean boosts.} Thus, we will consider a field \(\psi\) in a spin-\(\tfrac{1}{2}\) representation of rotations for \(d \geq 3\) or in an anyonic spin-\(s\) representation of rotations for \(d=2\). Then we have the Galilean generators
\be\label{eq:gal-rep-sch}
	J^{AB} = \begin{cases}
	\begin{pmatrix}
		0 & 0 \\
		0 & s \epsilon^{ab} 
	\end{pmatrix} \quad&\text{for}\quad  d = 2 \\[3ex]
	\begin{pmatrix}
		0 & 0 \\
		0 & \frac i 4 [ \gamma^a , \gamma^b ] 
	\end{pmatrix}\quad&\text{for}\quad  d \geq 3
	\end{cases}
\ee
and it can be verified that these satisfy the standard commutation relation of the Galilean algebra.

	If the mass of \(\psi\) is \(m\), the \(U(1)_{\mathbb{M}}\)-covariant derivative of \(\psi\) is then
\be
D_\mu \psi = \left(\partial_\mu - iqA_\mu - i m a_\mu - \frac i 2 \hat \omega_{\mu AB} J^{AB} \right)\psi
\ee
Here $q$ is the charge of the field \(\psi\) and \(A_\mu\) is a external electromagnetic field. However as discussed in \cite{GPR-fluids}, this derivative is not covariant under local Galilean boosts. The Galilean-covariant derivative acting on massive fields is given by
\be\label{eq:DI}
	D_I\psi \defn \begin{pmatrix} e^\mu_A D_\mu\psi, & i m\psi \end{pmatrix}
\ee

The Schr\"odinger action for such fields can then be written in a manifestly invariant form as \cite{GPR_geometry, GPR-fluids}
\begin{align}\label{schrodingerCovariant}
	\mc S = - \frac 1 {2m} \int d^{d+1}x | e | D_I \psi^\dagger D^I \psi ,
\end{align}
which one may check reduces to the standard Schr\"odinger action in flat spacetime after expanding in components. We are now in a position to perform a covariant calculation of the various currents defined in this note.

For this we will need the variation of the extended derivative operator acting on $\psi$. 
The non-covariant derivative $D_\mu$ simply varies with the mass gauge field $\delta D_\mu \psi = - i m \delta a_\mu \psi$, from which we find
\begin{gather}
	\delta (e^\mu_A D_\mu \psi) = - \delta e^B_\nu e^\nu_A e^\mu_B D_\mu \psi - i m \delta a_\nu e^\nu_A \psi = - \delta e^I_\nu e^\nu_A D_I \psi  .
\end{gather}
Including the variation of the spin connection and electromagnetic gauge field then gives
\begin{align}\label{derivVar}
	\delta D_I \psi = - \Pi^\mu{}_I \left( \delta e^J_\mu D_J \psi	+ \frac i 2 J^{AB} \psi \delta \omega_{\mu AB} + i q \psi \delta A_\mu \right).
\end{align}
Similarly, the variation of the volume element is
\begin{align}\label{volVar}
	\delta | e | = | e | e^\mu_A \delta e^A_\mu = | e | \Pi^\mu{}_I \delta e^I_\mu .
\end{align}

Using these, a straightforward computation gives the currents (on torsionless backgrounds)
\begin{subequations}\begin{align}
	s^{\mu AB} &= - \frac{i}{4m} \psi^\dagger J^{AB} \overset \leftrightarrow {\ms D}{}^\mu \psi ,\label{eq:Sch-spin-current} \\
	\tilde \tau^\mu{}_I &= - \frac{1}{2m} \left( {\ms D}^\mu \psi^\dagger D_I \psi + D_I \psi^\dagger {\ms D}^\mu \psi \right) + \frac 1 {2m} D_J \psi^\dagger D^J \psi \Pi^\mu{}_I ,  \\
	j^\mu &= - \frac{iq}{2m} \psi^\dagger \overset \leftrightarrow {\ms D} {}^\mu \psi
\end{align}\end{subequations}
where ${\ms D}^\mu \psi = e_A^\mu \ms D^A\psi$  is given by
\begin{align}\label{eq:funny-D}
	{\ms D}^A \psi = \Pi^{A I} D_I \psi = 
	\begin{pmatrix}
		im  \psi \\
		D^a \psi
	\end{pmatrix} .
\end{align}

Before writing the Cauchy currents, we note that the spin current \eqref{eq:Sch-spin-current} for the Schr\"odinger field is conserved on-shell i.e. \(D_\mu s^{\mu AB} = 0\) which can be shown as follows. Firstly, from \eqref{invariantTensors} we note that \(\Pi^{\mu I}\) is covariantly constant and so using \eqref{eq:Sch-spin-current} with \eqref{eq:funny-D} we have
\be\label{eq:Sch-spin-conserved}\begin{split}
	D_\mu s^{\mu AB} & = - \frac{i}{4m} \Pi^{\mu I} D_\mu \lb( \psi^\dagger J^{AB} \overset \leftrightarrow D_I \psi \rb) \\
		& = - \frac{i}{4m} D^I \lb( \psi^\dagger J^{AB} \overset \leftrightarrow D_I \psi \rb) \\
		& = - \frac{i}{4m} \lb( \psi^\dagger J^{AB} D^I D_I \psi \rb) + c.c. = 0
\end{split}\ee
where in the second line we have used the fact that \(\psi^\dagger J^{AB} \overset \leftrightarrow D_I \psi \) is \(U(1)_{\bb M}\) invariant which implies that \(\Pi^{\mu I}D_\mu = D^I\). Finally, the last line vanishes by the Schr\"odinger equation (on torsionless backgrounds) \(D^I D_I \psi = 0\).

The physical Cauchy currents are obtained from \eqref{stressMomentumImprovement} giving
\begin{subequations}\label{eq:cauchy_schrodinger}\begin{align}
	S^{\lambda \mu \nu} &= - \frac{i}{8m} \left( \psi^\dagger J^{\mu \nu} \overset \leftrightarrow {\ms D} {}^\lambda \psi  -  \psi^\dagger J^{\nu \lambda} \overset \leftrightarrow {\ms D} {}^\mu \psi  -  \psi^\dagger J^{\lambda \mu} \overset \leftrightarrow {\ms D} {}^\nu \psi \right) ,\\
	T^{\mu \nu} &=  \frac{1}{m}  {\ms D}^{(\mu} \psi^\dagger {\ms D}^{\nu)} \psi  - \frac 1 {2m} D_I \psi^\dagger D^I \psi h^{\mu \nu} - \frac{i}{2m} \nabla_\lambda \left( \psi^\dagger  J^{ \lambda (\mu} \overset \leftrightarrow {\ms D} {}^{\nu )}  \psi \right) ,\label{eq:Sch-Cauchy-stress-mass}\\
	\rho^\mu &= - \frac i 2 \psi^\dagger \overset \leftrightarrow {\ms D} {}^\mu \psi - \frac 1 2 \nabla_\nu \left( \psi^\dagger J^{\mu \nu} \psi \right) .
\end{align}\end{subequations}
where in \eqref{eq:Sch-Cauchy-stress-mass} we have used \eqref{eq:Sch-spin-conserved} to write the stress-mass tensor in a manifestly symmetric form as implied by the on-shell Ward identity for local Galilean transformations \eqref{eq:noether-gal-imp}.

We also give component expressions of the above equations for a spin-$\frac 1 2$ particle in flat, \(3+1\) dimensional spacetime. Let $S^i = \frac 1 2 \sigma^i$ be the Pauli spin operators, then the currents can be written as
\begin{subequations}\label{eq:schrodinger_component}\begin{align}
	s^{\mu}_j &= 
	\begin{pmatrix}
		\psi^\dagger S_j \psi \\
		- \frac{i}{2m} \psi^\dagger S_j \overset \leftrightarrow D{}^i \psi
	\end{pmatrix} , \\
	\tilde \varepsilon^\mu &=
	\begin{pmatrix}
		\frac{1}{2m} D_i \psi^\dagger D^i \psi \\
		- \frac 1 {2m} \left( D^i \psi^\dagger D_t \psi + D_t \psi^\dagger D^i \psi \right)
	\end{pmatrix}, \\
	T^{ij} & = \frac{1}{m} D^{(i} \psi^\dagger D^{j )} \psi + \left( \frac{i}{2} \psi^\dagger \overset \leftrightarrow  D_t \psi - \frac{1}{2m} D_i \psi^\dagger D^i \psi \right) \delta^{ij} - \frac i {2m} \epsilon^{kl(i} \partial_k \left( \psi^\dagger S_l \overset \leftrightarrow D{}^{j )} \psi \right) ,\\
	\rho^\mu &=
	\begin{pmatrix}
		m \psi^\dagger \psi \\
		- \frac{i}{2} \psi^\dagger \overset \leftrightarrow D{}^i \psi + \frac 1 2 \epsilon^{ijk} \partial_j \left( \psi^\dagger S_k \psi \right)
	\end{pmatrix} , \\
	j^\mu & =
	\begin{pmatrix}
		q \psi^\dagger \psi \\
		- \frac{iq}{2m} \psi^\dagger \overset \leftrightarrow D{}^i \psi
	\end{pmatrix}
\end{align}\end{subequations}
Note the energy current is roughly the anticommutator of the energy $- i D_t $ and velocity $- \frac{i} m D^i$ and represents kinetic energy being transported with the velocity of the particle. We have also written the spin current as $s^\mu_i =- \epsilon_{ijk} s^{jk \mu}=- \frac{i}{2m} \psi^\dagger S_i \overset \leftrightarrow {\ms D} {}^\mu \psi$ whose density is the spin density of standard quantum mechanics and whose current may also be interpreted along the lines of the energy current as it is half the anticommutator of the spin and velocity. By virtue of \eqref{eq:Sch-spin-conserved}, this spin current is conserved on-shell
\begin{align}
	\partial_\mu s^\mu_i = 0 .
\end{align}
The mass current \(\rho^\mu\) is $m$ times the probability current, plus a magnetization term $ \frac 1 2 \nabla \times \overset \rightarrow S$ that can be interpreted as the mass flow due to the non-uniform spin of matter. Note in particular that even in the single-constituent case in the presence of spinful matter, the charge and mass currents need not be aligned since the inhomogenous spin carries momentum. Finally, the stress tensor is the standard stress tensor for spinless Schr\"odinger fields, plus a contribution $\epsilon^{kl ( i} \partial_k s^{j)}_l$ arising from any non-uniform spin current. Of course, we do not display the momentum current since it is equal to the mass current on-shell.

\subsection{Non-relativistic Dirac field}\label{sec:Dirac}

Next we consider a massive field $\Psi$ that transforms non-trivially under local Galilean boosts in a \(3+1\)-dimensional spacetime. We note that, since Galilean boosts are non-compact they do not have finite-dimensional unitary representations and one cannot use the Schr\"odinger action \eqref{schrodingerCovariant}. 

We consider the spin-\(\tfrac{1}{2}\) representation\footnote{See \cite{Montigny,Niederle:2007xp} for higher spin representations.} originally discovered by L\'evy-Leblond \cite{LevyLeblond:1967zz}; see also section 3 of \cite{Montigny}.\footnote{To convert \cite{Montigny} to our conventions, take $J^{ab} = - \epsilon^{abc} (S_c)_{\mathrm{(MNN)}}$, $K^a = - \eta^a_{\mathrm{(MNN)}}$.} This representation involves a \(4\)-component field \(\Psi\) in which the action of the Galilean algebra generators is given by
\begin{align}
	J^{ab} = - \epsilon^{abc}
	\begin{pmatrix}
		\frac 1 2 \sigma_c & 0 \\
		0 & \frac 1 2 \sigma_c
	\end{pmatrix},
	&&K^{a} =
	\begin{pmatrix} 
		0 & 0 \\
		-\frac i 2 \sigma^a & 0
	\end{pmatrix}\label{eq:NR_dirac_rep}
\end{align}
Thus, \(\Psi\) contains two fields \(\psi\) and \(\chi\) that each transform as \(2\)-component spin-\(\tfrac{1}{2}\) fields under rotations and transform into each other under Galilean boosts i.e.
\be\label{eq:Dirac-to-Sch}
	\Psi = \begin{pmatrix}
			\psi \\ \chi
			\end{pmatrix}, \quad
	\Psi \mapsto -k_a K^a \Psi =
			\begin{pmatrix}
			\psi \\ \chi + \tfrac{i}{2} k_a\sigma^a \psi
			\end{pmatrix}
\ee
For such a representation with Galilean transformation $\Psi \rightarrow \Lambda_{\half} \Psi$, one can find a collection of matrices $\beta^I$ such that
\begin{align}\label{eq:beta-trans}
	\Lambda_{\half}^\dagger \beta^I \Lambda_{\half} = \Lambda^I{}_J \beta^J .
\end{align}
It can be verified that there is a matrix \(A\) which relates the \(\beta\)-matrices solving \eqref{eq:beta-trans} to the \(\gamma\)-matrices \(\gamma^I\) of the Lorentzian Clifford algebra in \(4+1\)-dimensions through a similarity transformation as
\be\label{eq:beta-to-gamma}
	\beta^0 = \tfrac{1}{2} A^{-1} \gamma^0 (\gamma^0 + \gamma^4) A, \quad \beta^a = A^{-1} \gamma^0\gamma^a A,\quad \beta^{\bb M} = - A^{-1} \gamma^0 (\gamma^0 - \gamma^4) A
\ee
If we take the Dirac representation for the $\gamma^I$ and $A = \begin{pmatrix}\half I & -\frac{i}{2}I \\ -\frac{i}{2}I & \half I\end{pmatrix}$ we retrieve the \(\beta\)-matrices found in equation (10) of \cite{Niederle:2007xp}, given by\footnote{Comparing to \cite{Niederle:2007xp}, we have $\beta^I = ( \beta_{0} , ~ \beta_a , - \beta_4)_{\mathrm{(NN)}}$.}
\begin{align}\label{betaMatrices}
	\beta^0 = 
	\begin{pmatrix}
		I & 0 \\
		0 & 0 
	\end{pmatrix},
	&&\beta^a = 
	\begin{pmatrix}
		0 & \sigma^a \\
		\sigma^a & 0 
	\end{pmatrix},
	&&\beta^\mathbb{M} =
	\begin{pmatrix}
		0 & 0 \\
		0 & - 2 I 
	\end{pmatrix}.
\end{align}\\

The non-relativistic Dirac action for \(\Psi\), on any curved spacetime, is then
\begin{align}\label{NRDirac}
	\mc S = \int d^4x | e |~ \frac i 2 \Psi^\dagger \beta^I \overset \leftrightarrow D_I \Psi,
\end{align}
Given the relations \eqref{eq:beta-to-gamma} the action \eqref{NRDirac} is the \(4+1\)-dimensional relativistic Dirac action written in terms of the extended representation, and the $c \rightarrow \infty$ limit of the \(3+1\)-dimensional relativistic Dirac action (see also \cite{Fuini:2015yva}).\\

From the action \eqref{NRDirac} we find the stress-energy and spin current as
\begin{align}
	\tilde \tau^\mu{}_I = \frac i 2 \Psi^\dagger \beta ^\mu \overset \leftrightarrow D_I \Psi - \frac i 2 \Psi^\dagger \beta ^J \overset \leftrightarrow D_J \Psi \Pi^\mu{}_I,
	&&s^{\mu AB} = \frac 1 4 \Psi^\dagger M^{\mu AB} \Psi ,
\end{align}
where we have defined $M^{C AB} = \beta^C J^{AB} + (J^{AB})^\dagger \beta ^C $ and as usual $\beta^C = \Pi^C{}_I \beta^I$. Computing the components of $M^{ABC}$ from \eqref{eq:NR_dirac_rep} and \eqref{betaMatrices}, we find that $M^{ABC}$ is in fact totally antisymmetric in its indices.

Further, using the relations in \eqref{eq:beta-to-gamma}, the total antisymmetry of $M^{CAB}$, and the equation of motion (on torsionless spacetimes) \(\beta^ID_I\Psi =0\) we find
\be\label{eq:Dirac-spin-conservation}
	D_\mu s^{\mu AB} = - \frac i 2 \Psi^\dagger \beta^{[A} \overset \leftrightarrow {\ms D} {}^{B]} \Psi
\ee\\

The physical currents in torsionless backgrounds are
\begin{subequations}\begin{align}
	S^{\lambda \mu \nu} & =- \frac 1 8 \Psi^\dagger M^{  \lambda \mu \nu } \Psi , \\
	T^{\mu \nu} & = - \frac i 2 \Psi^\dagger \beta ^{( \mu} \overset \leftrightarrow {\ms D} {}^{\nu ) } \Psi , \\ 
	\rho^\mu & =  \frac m 2  \Psi^\dagger \beta^\mu \Psi - \frac i 4 \Psi^\dagger \beta^0 \overset \leftrightarrow  {\ms D} {}^\mu \Psi, \\
	j^\mu & =  q  \Psi^\dagger \beta^\mu \Psi .
\end{align}\end{subequations}
In simplifying this we have used the equations of motion for \(\Psi\) as well as \eqref{eq:Dirac-spin-conservation}.\\

To cast the above currents in a more familiar form, we use the decompositon \eqref{eq:Dirac-to-Sch} to exapnd the action \eqref{NRDirac} as
\begin{align}
	\mc S = \int d^4 x | e | \left( \frac i 2 \psi^\dagger \overset \leftrightarrow D_0 \psi + \frac i 2 \psi^\dagger \sigma^a \overset \leftrightarrow D_a \chi + \frac i 2 \chi^\dagger \sigma^a \overset \leftrightarrow D_a \psi + 2 m \chi^\dagger \chi \right)
\end{align}
The bottom component \(\chi\) is auxiliary, satisfying the constraint (on torsionless spacetimes)
\begin{align}\label{auxiliarySol}
	\chi = - \frac{i}{2m} \sigma^a D_a \psi .
\end{align}

Plugging this in, on torsionless backgrounds and after integration by parts, gives
\begin{align}\label{gFactorAction}
	\mc S = \int d^4 x | e |  \left( \frac i 2 \psi^\dagger \overset \leftrightarrow D_0 \psi - \frac{\delta^{ab}}{2m} D_a \psi^\dagger D_b \psi + \frac q m B_a \psi^\dagger S^a\psi - \frac{R}{8m} \psi^\dagger \psi\right)
\end{align}
where $B^a =  \frac 1 2 \epsilon^{abc} F_{bc}$ is the magnetic field and $S^a = \frac 1 2 \sigma^a$ is the spin operator. \eqref{gFactorAction} is the Schr\"odinger action with a $g$-factor of 2 for the top spinor \(\psi\). The commutator of derivatives gives rise to the well-known $g$-factor coupling of the spin to the magnetic field, and also induces a non-minimal coupling to the Ricci scalar $R$.

	Using this we write the currents as
\begin{subequations}\begin{align}
	s^\mu_j &=
	\begin{pmatrix}
		\psi^\dagger S_j \psi \\
		- \frac{i}{2m} \psi^\dagger \sigma^k \overset \leftrightarrow D_k \psi \delta^i{}_j
	\end{pmatrix} ,
\nonumber \\
	\tilde \varepsilon^\mu &=
	\begin{pmatrix}
		- \frac{1}{4m} \psi^\dagger D_k D^k \psi - \frac{1}{4m} D_k D^k \psi^\dagger \psi - \frac q m B_i \psi^\dagger S^i \psi \\
		- \frac{1}{4m} D^i \psi^\dagger \overset \leftrightarrow D_t \psi + \frac{i}{2m} \epsilon^{ijk} D_j \psi^\dagger S_k \overset \leftrightarrow D_t \psi + c.c.
	\end{pmatrix} \\
	T^{ij} &= \frac{1}{2m} D^{(i} \psi^\dagger D^{j)} \psi - \frac 1{4m} \psi^\dagger D^{(i} D^{j)} \psi - \frac 1 {4m} D^{(i} D^{j)} \psi^\dagger \psi \nonumber\\
	&\quad + \frac q m B^{(i } \psi^\dagger S^{j)} \psi - \frac i{2m} \epsilon^{kl(i} \partial_k \left( \psi^\dagger S_l \overset \leftrightarrow D{}^{j)} \psi\right) \\
	\rho^\mu &=
	\begin{pmatrix}
		m \psi^\dagger \psi \\
		- \frac{i}{2} \psi^\dagger \overset \leftrightarrow D {}^i \psi + \frac 1 2\epsilon^{ijk} \partial_j \left( \psi^\dagger S_k \psi \right)
	\end{pmatrix} ,\\
	j^\mu &=
	\begin{pmatrix}
		q \psi^\dagger \psi \\
		- \frac{iq}{2m} \psi^\dagger \overset \leftrightarrow D {}^i \psi + \frac q m \epsilon^{ijk} \partial_j \left( \psi^\dagger S_k \psi \right)
	\end{pmatrix} .
\end{align}\end{subequations}
In particular, we have the standard charge and mass currents, plus magnetization currents arising from the magnetic moments that the $g$-factor attaches to particles. Note that the mass magnetization enters as though it had $g$-factor 1. In comparison to the Schr\"odinger case \eqref{eq:schrodinger_component}, these currents also have additional terms (modulo the equations of motion) in the energy current and stress arising from the non-minimal couplings to magnetic field and curvature found in \eqref{gFactorAction}.

\subsection{Wen-Zee term}

We conclude with the Wen-Zee term in \(2+1\)-dimensions, an important example from effective field theory for gapped systems in the presence of external curvature and electromagnetic field. The action for the Wen-Zee term is
\begin{align}\label{WenZee}
	\mc S = \int \frac{\kappa}{2 \pi} \df \omega \wedge d \df A .
\end{align}
This term famously encodes the Hall viscosity of quantum Hall systems \cite{Wen:1992ej}. Gauge invariance requires that the Wen-Zee coefficient $\kappa$ be integer valued. It then cannot be changed by a continuous deformation of the microscopic parameters of a system that does not close the gap and so characterizes topological phases of matter. Examination of quantum Hall effective actions with constraints from non-relativistic symmetries was initiated by Hoyos and Son in \cite{Hoyos:2011ez}. In this work the authors also impose additional symmetries owing to the single component nature of the quantum Hall fluid (see \cite{Geracie:2016inm} for a manifestly covariant way of implementing these symmetries). In this section we will reproduce known results for the stress and energy current induced by the Wen-Zee term as a simple example in the use of the formalism given above (also see \cite{Bradlyn:2014wla} for a similar computation). For those interested in a full general effective action the results can be found in \cite{GeracieThesis}. It would be interesting to examine, which, if any, of these terms require the introduction of gapless edge modes, along the lines of \cite{Gromov:2015fda}.\\

In the original treatment of \cite{Wen:1992ej}, the spin connection $\df \omega$ appearing in the Wen-Zee term is the torsionfree connection purely for spatial rotations defined in terms of the spatial coframes. However to preserve Galilean invariance we use the full spacetime spin connection $\hat {\df \omega}_{AB}$ . In \(2+1\)-dimensions we can covariantly extract the spatial part
\begin{align}
	\df \omega = \frac 1 2 \hat {\df \omega}_{AB} \epsilon^{ABC} n_C
\end{align}
which reduces to the one used by \cite{Wen:1992ej} when restricted to time-independent curved geometries. It is this $\df \omega$ that appears in (\ref{WenZee}).

The induced charge current is straightforward to calculate
\begin{align}
	j^\mu = \frac{\kappa}{4 \pi} \varepsilon^{\mu \nu \lambda} R_{\nu \lambda} 
\end{align}
where $R_{\mu\nu} \equiv \df R = d \df \omega$ is the curvature \(2\)-form. This in particular gives the well-known result that the Wen-Zee term attaches charge to Ricci curvature $R = \frac 1 2 \varepsilon^{\mu \nu} R_{\mu \nu}$
\begin{align}
	j^0 = \frac{\kappa R}{2 \pi} .
\end{align}

The stress-energy vanishes since it is defined at fixed connection
\begin{align}
	\tilde \tau^\mu{}_I = 0
\end{align}
so that in particular we see that the Wen-Zee term makes no contribution to the physical energy density or current, as in \cite{Bradlyn:2014wla}. The spin current is
\begin{align}
	s^{\mu AB} = \frac{\kappa}{ 8 \pi} \varepsilon^{\mu \nu \lambda} F_{\nu \lambda} \epsilon^{AB} = \frac{\kappa B}{4 \pi} u^\mu \epsilon^{AB} ,
\end{align}
where we have introduced the covariant drift 3-velocity
\begin{align}
	u^A = \frac{1}{2B}\varepsilon^{ABC} F_{BC}
		= \begin{pmatrix}
				1 \\
				\frac{\epsilon^{ab} E_b}{B}
			\end{pmatrix}.
\end{align}
Note that, due to the the Bianchi identity $d \df F = 0 \implies \nabla_\mu ( B u^\mu ) = 0$, the spin current is  identically conserved \(D_\mu s^{\mu AB} = 0\).

The Cauchy stress-mass is then
\begin{align}
	T^{\mu \nu} &= -2 \nabla_\lambda S^{\nu \mu \lambda} \nonumber \\
		&= \frac{\kappa B}{2 \pi} \varepsilon^{\lambda ( \mu} \nabla_\lambda u^{\nu )}
- \frac{\kappa}{2 \pi} u^{( \mu} \varepsilon^{\nu ) \lambda} \nabla_\lambda B .
\end{align}
where we have discarded a term due to the conservation of the spin current yielding a manifestly symmetric stress-mass tensor.\footnote{Note that the stress-mass tensor needs to be identically symmetric, not simply symmetric on-shell, since we have integrated out the matter fields.} The first term is the standard Hall viscosity term, while the final term gives a mass magnetization current in the presence of inhomogeneities in the external magnetic field
\begin{align}
	\rho^\mu = - \frac{\kappa}{4 \pi} \varepsilon^{\mu \nu} \nabla_\nu B .
\end{align}

\section{Conclusions}
Working with a manifestly covariant geometric description, given by Bargmann spacetimes, we define the physical energy current, stress tensor, and mass current for any Galilean invariant physical system with spin. We find that when the stress, mass and momentum are appropriately defined, the stress tensor is symmetric, and momentum and mass currents coincide as a consequence of manifest local Galilean invariance. We also argue that the physical energy current is naturally defined via variation at fixed connection, not fixed torsion. 

	While we have worked out some illustrative examples, it would be of interest to use this formalism to extend the analysis of \cite{Geracie:2014zha,Jensen:2014aia,GPR-fluids} to spinful fluids, and that of \cite{Geracie:2016inm} to spinful electrons. One could also investigate non-relativistic scale anomalies for spinful non-relativistic fields following \cite{Baggio:2011ha, Jensen:2014hqa, Arav:2014goa, Arav:2016xjc, Pal:2016rpz}. Another potentially interesting application would be to examine the linear response in Son's Dirac theory of the half-filled Landau-level \cite{Son:2015xqa}.

\section*{Acknowledgments}
This work is supported in part by the NSF grants DMR-MRSEC 1420709, PHY~12-02718 and PHY~15-05124. M.G. is supported in part by the University of California. K.P. is supported in part by the NSF grant PHY-1404105. M.M.R. is supported in part by the DOE grant DE-FG02-13ER41958.

\appendix
\section{Symmetries of the Riemann tensor}\label{app:Riemann}
In the main text, we required the symmetries of the Newton-Cartan Riemann tensor to derive equation (\ref{cauchyEquation}). These identities involve a few subtleties not present in the pseudo-Riemannian case, so we collect their derivations here. Since we are interested in the Ward identities on unrestricted Bargmann geometries, we will present these symmetries on spacetimes with general extended torsion $\df T^I$ (the torsionless case can be found in \cite{Mal-book}). They are
\begin{subequations}\label{id4}\begin{align}
	\hat R_{(AB)\mu \nu} &= \hat R_{AB(\mu \nu)}= 0,\label{id1} \\
	R^A{}_{[BCD]} &= \frac 1 3 ( D T^A )_{BCD}, \label{id2}  \\
	R_{IJKL} &= R_{KLIJ} + \frac 1 2 \left( (D T_I )_{JKL} + ( D T_J )_{KIL} + ( D T_K )_{ILJ} + (D T_L )_{IJK} \right)\label{id3} ,  \\
	D_{[\mu|} \hat R_{AB| \nu \lambda ]} &= T^\rho{}_{[\mu \nu|} \hat R_{AB| \lambda ] \rho} .
\end{align}\end{subequations}
Where we have defined $\hat R_{ABCD}$ as the unique object anti-symmetric in it's first two indices such that $R^A{}_{BCD} = h^{AE}\hat R_{EBCD}$. Equivalently
\begin{align}
	\hat {\df R}_{AB} = d \hat {\df \omega}_{AB} + \hat {\df \omega}_{AC} \wedge \hat {\df \omega} {}^C{}_B .
\end{align}
The derivation of (\ref{cauchyEquation}) requires only the first three of these identities, but we include the Bianchi identity for completeness. Contracting equation (\ref{id2}) with $\delta^B{}_A$, we also find
\begin{align}
	2 R_{[ \mu \nu]} = 3 \nabla_{[\mu} T^\lambda{}_{\lambda \nu]} + T^\lambda{}_{\lambda \rho} T^\rho{}_{\mu \nu},
\end{align}
which was used in obtaining equation (\ref{cauchyIntermediate}).

The first identity follows trivially from the definition of $\df {\hat R}_{AB}$ while the derivations of (\ref{id2}) and (\ref{id4}) from
\begin{align}
	D \df T^A = \df R^A{}_B \wedge \df e^B,
	&&\text{and}
	&&D \df {\hat R}_{AB} = 0
\end{align}
are identical to the pseudo-Riemannian case. The only identity that requires some care is (\ref{id3}), which is most easily stated when valued in the extended representation. By $R^I{}_{JKL}$ we mean the curvature two-form, valued in the extended representation of $\mathfrak{gal}(d)$, with spacetime indices pulled back to the extended representation using the Galilean invariant projector 
\begin{align}
	R^I{}_{JKL} = ( R^I{}_J )_{\mu \nu} \Pi^\mu{}_I \Pi^\nu{}_J .
\end{align}
One can write (\ref{id3}) in terms of $\hat R_{ABCD}$ using the three-form $\hat{ \df \Xi}_A$ introduced in (\ref{TAdef}). Since all indices in this equation are $n^I$ orthogonal, it is simply the pullback of an equation valued in the covector representation of $Gal(d)$
\begin{gather}
	\hat R_{ABCD} = \hat R_{CDAB} + \frac 1 2 \left( \hat{  \Xi}_{ABCD} +  \hat{  \Xi}_{BCAD} + \hat{  \Xi}_{CADB} + \hat{  \Xi}_{DABC} \right) ,
\end{gather}
where we have used $\df R_{IJ} = \df {\hat R}_{AB} \Pi^A{}_I \Pi^B{}_J$ and $D \df T_I = \Pi^A{}_I \hat{ \df \Xi}_A$.
However, the proof of this identity is most naturally carried out in it's extended form. 

To prove (\ref{id3}) we begin with
\begin{align}
	D \df T^I = \df R^I{}_J \wedge \df e^J ,
\end{align}
which written in tensor notation reads
\begin{align}
	(R_{IJ})_{[\mu \nu} e^J_{\lambda ]} = \frac 1 3 (D T_I)_{\mu \nu \lambda}.
\end{align}
Now let us pull this back to an equation involving only extended indices using $\Pi^\mu{}_I$. One may check by an explicit computation in components that
\begin{align}
	e^I_\mu \Pi^\mu{}_J = \delta^I{}_J + n^I a_J
\end{align}
where $a_I = ( a_A, -1 )$ --- one can check that $a_I$ indeed transforms covariantly as indicated by its index structure. Since $\df R_{IJ} n^J = 0$, the second term drops out and we find
\begin{align}
	R_{I[JKL]} = \frac 1 3 ( D T_I )_{JKL} .
\end{align}
This is simply the extended index version of (\ref{id2}), which one can obtain from here by noting that both sides are $n^I$ orthogonal in all their indices. (\ref{id3}) then follows exactly as in the pseudo-Riemannian case by repeated applications of this equation along with
\begin{align}
	R_{(IJ)KL} = R_{IJ(KL)} = 0 .
\end{align}


\bibliographystyle{JHEP}
\bibliography{WenZeerefs}

\end{document}